\documentclass[letterpaper, 10 pt, conference]{ieeeconf}  % Comment this line out if you need a4paper

\IEEEoverridecommandlockouts           % This command is only needed if % you want to use the \thanks command
                                                          
\usepackage{amsthm}
\theoremstyle{definition}
\newtheorem{definition}{Definition}[]
\newtheorem{remark}{Remark}

\usepackage{multirow}
\usepackage{lettrine}
\usepackage{caption} 
\usepackage{graphicx,float}
\usepackage{algorithm}
\usepackage{algpseudocode}
\algnewcommand{\Initialize}[1]{%
  \State \textbf{Initialize:}
  \Statex \hspace*{\algorithmicindent}\parbox[t]{1\linewidth}{\raggedright #1}
}

\usepackage{tikz}
\usetikzlibrary{automata, positioning, arrows.meta,calc,cd}
\usetikzlibrary{patterns}
\usetikzlibrary{datavisualization}

\tikzset{
->, % makes the edges directed
>=stealth, % makes the arrow heads bold
node distance=3cm, % specifies the minimum distance between two nodes. Change if necessary.
every state/.style={thick, fill=gray!10}, % sets the properties for each ’state’ node
initial text=$ $, % sets the text that appears on the start arrow
}
\usepackage[fleqn]{amsmath}
\usepackage{amssymb}
\usepackage{cite}
\usepackage{lettrine}
\let\labelindent\relax
\usepackage{enumitem}
\usepackage{multicol}
\usepackage{pgfplots}
\makeatletter
\newenvironment{customlegend}[1][]{%
    \begingroup
    \pgfplots@init@cleared@structures
    \pgfplotsset{#1}%
}{%
    \pgfplots@createlegend
    \endgroup
}%

\def\addlegendimage{\pgfplots@addlegendimage}
\makeatother
\usepackage{subfigure}
\usepackage[hang,flushmargin]{footmisc}
\newcommand{\algorithmfootnote}[2][\footnotesize]{%
  \let\old@algocf@finish\@algocf@finish% Store algorithm finish macro
  \def\@algocf@finish{\old@algocf@finish% Update finish macro to insert "footnote"
    \leavevmode\rlap{\begin{minipage}{\linewidth}
    #1#2
    \end{minipage}}%
  }%
}
\usepackage[T1]{fontenc}
\usepackage[export]{adjustbox}
\usepackage{hyperref}
\usepackage{orcidlink}
\usepackage{caption}
\usepackage{booktabs,makecell}

\title{Risk-Averse Model Predictive Control for Priced Timed Automata}
\date{June 2022}
\author{Mostafa Tavakkoli Anbarani\textsuperscript{1$\ast$} \thanks{\textsuperscript{1} Department of Mechanical Engineering, Pennsylvania State University, PA, USA\hspace{0.1 cm}mkt5457@psu.edu}, Efe C. Balta\textsuperscript{2} \thanks{\textsuperscript{2} Automatic Control Laboratory, ETH Z\"urich, Z\"urich, Switzerland\hspace{0.1 cm}ebalta@ethz.ch}, Rômulo Meira-Góes\textsuperscript{3} \thanks{\textsuperscript{3} Department of Electrical Engineering, Pennsylvania State University, PA, USA\hspace{0.1 cm}romulo@psu.edu}, and Ilya Kovalenko\textsuperscript{4} \thanks{\textsuperscript{4} Department of Mechanical Engineering, Pennsylvania State University, PA, USA\hspace{0.2 cm}iqk5135@psu.edu}}
\begin{document}

\maketitle
% \textbf{\textit{Abstract}- We have developed a Risk-Averse Priced Timed Automata (PTA) Model Predictive Control (MPC) scheme to increase system flexibility. In this paper, we developed risk-averse PTA MPC scheme using the PTA MPC formalism. Further, we introduced risk-averse PTA MPC multi-objective optimization problem to minimize the cost and risk, simultaneously. An example from manufacturing systems is presented to show the application of proposed controller theme.}
\textbf{\textit{Abstract}- In this paper, we propose a Risk-Averse Priced Timed Automata (PTA) Model Predictive Control (MPC) framework to increase flexibility of cyber-physical systems. 
To improve flexibility in these systems, our risk-averse framework solves a multi-objective optimization problem to minimize the cost and risk, simultaneously.
While minimizing cost ensures the least effort to achieve a task, minimizing risk provides guarantees on the feasibility of the task even during uncertainty.
Our framework explores the trade-off between these two qualities to obtain risk-averse control actions. The solution of risk-averse PTA MPC dynamic decision-making algorithm reacts relatively better to PTA changes compared to PTA MPC without risk-averse feature.
An example from manufacturing systems is presented to show the application of the proposed control strategy.}
\par
\textbf{Keywords- Model Predictive Control, Risk-Averse, Flexibility, Priced Timed Automaton.}

\section{Introduction}
\label{section-intro}
\lettrine[findent=2pt]{\textbf{A}}{ }Cyber Physical System (CPS) is an automated system that integrates physical reality with communication networks and computing infrastructures. \cite{wang_current_2015,acatech_-_national_academy_of_science_and_engineering_2011_cyber-physical_2011}. Subsystems in a CPS are often controlled or monitored by a computer-based algorithm utilizing high level of automation and situation-awareness. This requires the integration of a large amount of data and the use of complex algorithms to handle the data. For efficient performance, CPSs will often require a control architecture that can robustly coordinate various subsystem functions under unpredictable system dynamics or performance requirement changes. %and that can perform robustly under disturbances.
\par
One formalism to model CPS is as Discrete-Event System (DES). In DES, the state space of the system is described by a discrete set and state transitions \cite{cassandras_introduction_2021}. DES modeling method is used specially for systems where the control signal is scheduled with respect to \textit{events} rather than time. To describe performance metrics for DES in terms of temporal variables, Timed Automata (TA) have been previously used to model CPS behaviour \cite{behrmann2005optimal,kovalenko_priced_2020, cassandras_sensitivity_2008}.
TAs are a type of finite-state machine (FSM) where time-based dynamics and constraints of a system are described by adding a set of clocks to the model.
TAs have been extended to Priced Timed Automata (PTA) to capture costs associated with staying at a specific state  \cite{balta_model_2022,kovalenko_cooperative_2022,hutchison_priced_2005}. Formulating these temporal cost enables the formation of an optimization problem which can be solved using automata model checkers or theorem provers \cite{behrmann_scheduling_2005}. PTA applications have been used to control various systems, including bio-manufacturing \cite{hekmatnejad_task_2019}, airport traffic control \cite{goos_as_2001}, Cyber-Physical Manufacturing (CPM) \cite{tilbury_cyber-physical_2019,panetto_challenges_2019}, and  smart process planning \cite{gray2020industrial,ionescu2020digital}. 
% \par
% PTAs are models that use distinct state locations and a set of continuous clocks comprise the system states. 
% States and edges are associated with linear cost rates and cost increment, respectively \cite{balta_model_2022}.
\par
For PTA, flexibility is defined as maintaining the ability to reach the desired set of states while being subjected to automata uncertainties \cite{balta_model_2022,kovalenko_cooperative_2022}. This includes robustness to temporary or permanent state removal, changes in temporal cost associated with each state, and changes in the desired state set. 
\par
One way to increase system flexibility in face of model uncertainty is to use Model-Predictive-Control (MPC). MPC is a control scheme which uses a dynamically updated prediction model to generate the control law. The control law is implicitly formulated based on the prediction model to forecast system behavior, and optimize the actions to operate under the sequence of states which minimizes the cost objective \cite{tarragona_systematic_2021,rawlings_model_2009,schwenzer_review_2021}. Recently, MPC has been developed for PTA \cite{kovalenko_priced_2020}. However, MPC with PTA formulation often represents single objective constrained cost optimization and neglects risks related to model uncertainty. Therefore, the optimal solution found by PTA MPC is prone to execution failure as a result of disturbances that stop the execution of the planned path sequence \cite{schwenzer_review_2021-1,noauthor_tutorial_2000,li_resource_2023}. For example, a traveler flight itinerary represents a PTA, where the goal is to reach the destination (desired state set), in a series of flights (edges), while minimizing cost (cost objective). Missing a connection flight (state failure), can cause a failure to reach the destination (failure in reaching the desired state set).
Existing literature introduced MPC for systems described as PTA and perform constrained cost optimization. However, to the best of the authors knowledge, no previous work studied a risk-averse PTA-MPC minimizing both risk and cost, simultaneously. 
\par
In this work, we propose a framework  for a risk-averse MPC that uses existing an PTA modeling formalism for CPSs. Specifically, this work describes an MPC architecture that utilizes PTA models to encode time-based constraints and costs in order to solve the multi-objective optimization problem of cost and risk-averseness. Behavior of a risk-averse MPCs that use PTA-based model is compared to the behavior of existing MPC with PTAs described in \cite{balta_model_2022} using a simulation.
\par
The rest of this paper is organized as follows. Section \ref{sec-modeling} describes PTA modeling. In Section \ref{sec-PTA MPC} the risk-averse PTA MPC scheme is discussed. Section \ref{sec-case study}, shows several manufacturing case studies. Finally, concluding remarks and future works are given in Section \ref{sec-conclusion}.

\section{PTA Modeling}
\label{sec-modeling}
This section introduces definitions needed for describing the proposed risk-averse PTA MPC framework.
\begin{definition}[Priced Timed Automata]
A priced timed automaton (PTA) is defined as follows:
\begin{equation}
    \mathcal{A}=(Q, C,\Sigma, E, I, R, P, q_{0})
\end{equation}
where $Q=\{q^{1},q^{2},\cdots,q^{n_{q}}\}$ is a finite set of state locations, $C=c^{1}\times c^{2}\times \cdots c^{n_{c}}=\mathbb{R} ^{n_{c}}_{\geq 0}$ is the clock set space, $\Sigma=\langle\sigma^{1},\cdots,\sigma^{n_{\sigma}}\rangle$ is an ordered finite set of desired states, $E\subseteq Q \times \mathcal{B}(C) \times \Sigma \times Q$ is a finite set of edges, $I: Q \to \mathcal{B} (C)$ is the invariant operator, $R: E \times C \to C$ is the reset operator, $P: Q \cup E \to [0,\infty)$ maps locations and edges to costs, and $q_{0} \in Q$ is the initial location.
\end{definition}

% \begin{definition}[Connectivity]
% \label{def-connectivity}
% Given automaton $\mathcal{A}=(Q, C,\allowbreak \Sigma, E, I, R, P, q_{0})$, initial state $q_{0}$ and state $q^{m}$ are said to be connected if there exists a path from $q_{0}$ to $q^{m}$. Otherwise, $q^{m}$ is disconnected.
% \par
% If all states of $\mathcal{A}$ are connected, $\mathcal{A}$ is said to be connected. Otherwise, $\mathcal{A}$ is disconnected.
% \end{definition}
In this work, the automaton defined by user as input to model for MPC is labeled as "Total Automaton" which itself splits into two other automata; "Original Automaton" and "Redundant Automaton". The total automaton is prone to risk of failure, a situation in which some of states are excluded from the PTA. The part of total automaton which describes the system PTA under ideal case where no failure exists is called original automaton, whereas the redundant automaton describes an automaton which is enabled under failures.
%To capture possible changes in the environment, e.g., state failure, extra path availability, etc., we define three different automata: Original, Redundant and Total.
%The original automaton captures the system with basic operations, whereas the redundant automaton captures possible extra paths. 
%Lastly, the union of the two is the total automaton which is defined by the user as the input model for MPC.
\begin{remark}
In this paper $\langle.\rangle$ and $\{.\}$ denote ordered sets and sets, respectively. For an ordered set, the order at which array elements occur is important. Therefore, two ordered sets are equal if they have the same elements with all elements occurring in the same order.
\end{remark}
\begin{definition}[Total, Original, and Redundant Automata]
Given the system Total Automaton defined by the user as $\mathcal{A}^{T}=(Q, C,\Sigma, E, I, R, P, q_{0})$, the Original Automaton $\mathcal{A}^{O} \subseteq \mathcal{A}^{T}$ is defined by $\mathcal{A}^{O}=(Q^{O}, C^{O},\Sigma, E^{O}, I^{O}, \allowbreak R^{O}, P^{O}, q_{0})$. Moreover, Redundant Automaton $\mathcal{A}^{R}=\cup_{v=1}^{k}\mathcal{A}^{R_{v}}$ is defined by $\mathcal{A}^{R}=\cup_{v=1}^{k}(Q^{R_{v}}, C^{R_{v}},\Sigma, E^{R_{v}}, I^{R_{v}}, R^{R_{v}},\allowbreak P^{R_{v}}, q_{R_{v}})$,  where $k$ is the number of redundant paths, $q_{R_{v}}$ and $Q^{R_{v}}$ are the redundant initial state and the set of redundant state locations defined as below:
\begin{align}
    & Q^{R_{v}}=\{ \langle q^{i},q^{r_{1}},\cdots,q^{r_{n}},q^{j}\rangle | q^{i},q^{j} \in Q^{O}, \\
    &\qquad \quad \forall s \in \{r_{1},\cdots,r_{n}\}, \hspace{0.2cm} q^{s}\in Q^{T}-Q^{O} \} \notag\\
    & q_{R_{v}}=q^{i}
\end{align}
Thus, $Q^{R}=\cup_{v=1}^{k}Q^{R_{v}}$ represents an ordered state set in which only the first and last states of all redundant paths exist in the original system automaton. The rest of elements of $\mathcal{A}^{R}$ are user-defined. 
\end{definition}
The following definitions determine the properties that two paths in the original automaton $A^{O}$ should share so that they can be connected by a redundant path from redundant automaton $\mathcal{A}^{R}$ in case of state failure in the path the redundant path is branching from. 
\begin{definition}[End Parity]
The ordered set $EP_{\Sigma}$ is defined as the end parity of ordered set $\Sigma=\langle q^{0},q^{1},\cdots,q^{n} \rangle$ if:
\begin{equation}
    \exists q^{k}\in \Sigma: EP_{\Sigma}=\langle q^{k},\cdots,q^{n} \rangle
\end{equation}
\end{definition}

\begin{definition}[Equivalent Paths]
Given the original system automaton $\mathcal{A}^{O}$ where $EP_{\Sigma}$ is the end parity of the set of desired states $\Sigma=\langle \sigma_{1}, \cdots, \sigma_{n} \rangle$, and redundant path $Q^{R}=\langle q^{i},q^{r_{1}},\cdots,q^{r_{n}},q^{j}\rangle$ starting at state $q^{i}$ and ending in state $q^{j}$, then two paths of arbitrary length $U_{1}=\langle \cdots ,q^{i-1},q^{i},q^{i+1},\cdots \rangle$ and $U_{2}=\langle \cdots ,q^{j-1},q^{j},q^{j+1},\cdots \rangle$ are defined equivalent paths if their two subsets $U_{s1}=\langle q^{i},q^{i+1},\cdots \rangle$ and $U_{s2}=\langle q^{j},q^{j+1},\cdots \rangle$ have the following properties:
\begin{align}
&U_{1}\subseteq\mathcal{L}(\mathcal{A}^{O}),U_{2}\subseteq\mathcal{L}(\mathcal{A}^{O})\\
&EP_{\Sigma} \in U_{s1},U_{s2}
\end{align}
where $\mathcal{L}(\mathcal{A}^{O})$ is the set of all feasible paths in $\mathcal{A}^{O}$ \cite{balta_model_2022}. As per above, $U_{1}$ and $U_{2}$ are equivalent paths if the share the same end parity of desired state set.
\end{definition}

The following definition explains the state failure and the resulting path failure.
\begin{definition}[Legal State/Path]
% Given the original automaton tuple $\mathcal{A}=(Q^{O}, C^{O},\Sigma, E^{O}, I^{O}, R^{O}, P^{O}, q_{0})$, a state $q^{i} \in Q^{O}$ is legal if it has connectivity with respect to $q_{0}$. 
Given the original automaton tuple $\mathcal{A}^O=(Q^{O}, C^{O},\Sigma, E^{O}, I^{O}, R^{O}, P^{O}, q_{0})$, a state $q^{i} \in Q^{O}$ is legal if there exists a path between $q^i$ and the initial state $q_0$. 
Otherwise $q^{i}$ is called illegal and it is referred to as \textit{state failure}. 
Similarly, a path $\beta=\langle q^{k}, \cdots, q^{k+l}  \rangle \in \mathcal{A}^{O}$ is called legal if all states in the path are legal. 
Otherwise the path $\beta$ is called illegal path and the condition is referred to as \textit{path failure}.
\end{definition}
\begin{remark}
In this work, the general term \textit{failure} refers to both the state failure and path failure.
\end{remark}
The following definition introduces the concept of Emergency Declaration Signal (EDS) that has two objectives: (1) to indicate the availability of a redundant path $Q^{R}$; (2) to determine if the states of equivalent path past to the one redundant path is starting from $U_{s1}$ are still legal.
The EDS shows three status cases: Case Normal, Case Auxiliary, and Case Emergency. These cases identify if a failure occurred as well as the type of failure. 
% As will be discussed later, under Case Normal, $Q^{R}$ is unavailable and path $U_{s1}$ is legal. 
Case Normal captures the situation when a failure has not occurred.
In this scenario, the states in $Q^{R}$ are not unavailable and the path $U_{s1}$ is legal.
Case Auxiliary also captures the situation of no failure, but it makes states in $Q^{R}$ available and the path $U_{s1}$ is legal.
Lastly, Case Emergency happens when a failure has occurred.
In this scenario, the states in $Q^R$ are available and the path $U_{s1}$ is illegal.
\par
\begin{figure}[t]
    \begin{center}
    \begin{adjustbox}{minipage=\textwidth,scale=0.65}
    \vfil
    \subfigure[\label{figure-original language}]{
    \begin{tikzpicture}[on grid, state/.style={thick, fill=gray!10,circle, draw, minimum size=1cm},>={Stealth[inset=0pt,length=6pt,angle'=28,round]}]

    \node[state] at(0,0) (q1) {$q^{i-1}$};
    \node[state] (q2) at(2,0) {$q^{i}$};
    \node[state] (q3) at(4,0) {$q^{i+1}$};
    \node[state] (q6) at(6,-2) {$q^{j}$};
    \node[state] (q7) at(4,-2) {$q^{j-1}$};
    \node[state] (q8) at(8,-2) {$q^{j+1}$};
    \node[state, draw=none, fill=none] (qdots1) at(-2,0) {$\boldsymbol{\cdots}$};
    \node[state, draw=none, fill=none] (qdots2) at(6,0) {$\boldsymbol{\cdots}$};
    \node[state, draw=none, fill=none] (qdots6) at(2,-2) {$\boldsymbol{\cdots}$};
    \node[state, draw=none, fill=none] (qdots7) at(10,-2) {$\boldsymbol{\cdots}$};
    \draw
    (q3) -- (qdots2)
    (qdots1) edge node []{} (q1)
    (q1) edge node [below left]{} (q2)
    (q7) edge node [below] {} (q6)
    (q6) edge node [below] {} (q8)
    (qdots6) edge node [below] {} (q7)
    (q8) edge node [below] {} (qdots7)
    (q2) edge node [below] {} (q3);
    
    (q3) edge node []{} (qdots2)
    \end{tikzpicture}
    }
    \vfil
    \subfigure[\label{figure-auxiliary language}]{
    \begin{tikzpicture}[on grid, state/.style={thick, fill=gray!10,circle, draw, minimum size=1cm},>={Stealth[inset=0pt,length=6pt,angle'=28,round]}]

    \node[state] at(0,0) (q1) {$q^{i-1}$};
    \node[state] (q2) at(2,0) {$q^{i}$};
    \node[state] (q3) at(4,0) {$q^{i+1}$};
    \node[state, fill=red!20, densely dashed] (q4) at(3,-1) {$q^{r_{1}}$};
    \node[state, fill=red!20, densely dashed] (q5) at(7,-1) {$q^{r_{n}}$};
    \node[state] (q6) at(6,-2) {$q^{j}$};
    \node[state] (q7) at(4,-2) {$q^{j-1}$};
    \node[state] (q8) at(8,-2) {$q^{j+1}$};
    \node[state, draw=none, fill=none] (qdots3) at(-2,0) {$\boldsymbol{\cdots}$};
    \node[state, draw=none, fill=none] (qdots4) at(6,0) {$\boldsymbol{\cdots}$};
    \node[state, draw=none, fill=none] (qdots5) at(5,-1) {$\boldsymbol{\cdots}$};
    \node[state, draw=none, fill=none] (qdots6) at(2,-2) {$\boldsymbol{\cdots}$};
    \node[state, draw=none, fill=none] (qdots7) at(10,-2) {$\boldsymbol{\cdots}$};

    \draw
    (q3) -- (qdots4)
    (qdots3) edge node []{} (q1)
    (q1) edge node [below left]{} (q2)
    (q2) edge node [below] {} (q3)
    (q3) edge node []{} (qdots4)
    (q2) edge [bend right,densely dashed] node [] {} (q4)
    (q4) edge [densely dashed] node [] {} (qdots5)
    (q5) edge [bend left,densely dashed] node []{} (q6)
    (q7) edge node [below] {} (q6)
    (q6) edge node [below] {} (q8)
    (qdots6) edge node []{} (q7)
    (q8) edge node []{} (qdots7)
    (qdots5) edge [densely dashed] node []{} (q5);
    \end{tikzpicture}
    }
    \vfil
    \subfigure[\label{figure-emergency language}]{
    \begin{tikzpicture}[on grid, state/.style={thick, fill=gray!10,circle, draw, minimum size=1cm},>={Stealth[inset=0pt,length=6pt,angle'=28,round]}]

    \node[state] at(0,0) (q1) {$q^{i-1}$};
    \node[state] (q2) at(2,0) {$q^{i}$};
    \node[state, fill=red!20, densely dashed] (q4) at(3,-1) {$q^{r_{1}}$};
    \node[state, fill=red!20, densely dashed] (q5) at(7,-1) {$q^{r_{n}}$};
    \node[state] (q6) at(6,-2) {$q^{j}$};
    \node[state] (q7) at(4,-2) {$q^{j-1}$};
    \node[state] (q8) at(8,-2) {$q^{j+1}$};
    \node[state, draw=none, fill=none] (qdots3) at(-2,0) {$\boldsymbol{\cdots}$};
    \node[state, draw=none, fill=none] (qdots5) at(5,-1) {$\boldsymbol{\cdots}$};
    \node[state, draw=none, fill=none] (qdots6) at(2,-2) {$\boldsymbol{\cdots}$};
    \node[state, draw=none, fill=none] (qdots7) at(10,-2) {$\boldsymbol{\cdots}$};
    
    \draw
    (qdots3) edge node []{} (q1)
    (q1) edge node [below left]{} (q2)
    (q2) edge [bend right,densely dashed] node [] {} (q4)
    (q4) edge [densely dashed] node [] {} (qdots5)
    (q5) edge [bend left,densely dashed] node []{} (q6)
    (q7) edge node [below] {} (q6)
    (q6) edge node [below] {} (q8)
    (qdots6) edge node []{} (q7)
    (q8) edge node []{} (qdots7)
    (qdots5) edge [densely dashed] node []{} (q5);
    \end{tikzpicture}
    }
    \end{adjustbox}
    \end{center}
    \caption{PTA of two equivalent paths (solid-grey), with redundant path (dashed-red) for: \subref{figure-original language} Case Normal ($z_{i}=0$), \subref{figure-auxiliary language} Case Auxiliary ($z_{i}=-1$) and; \subref{figure-emergency language} Case Emergency ($z_{i}=1$).}
    \label{figure-original,aux and emergency paths}
\end{figure}
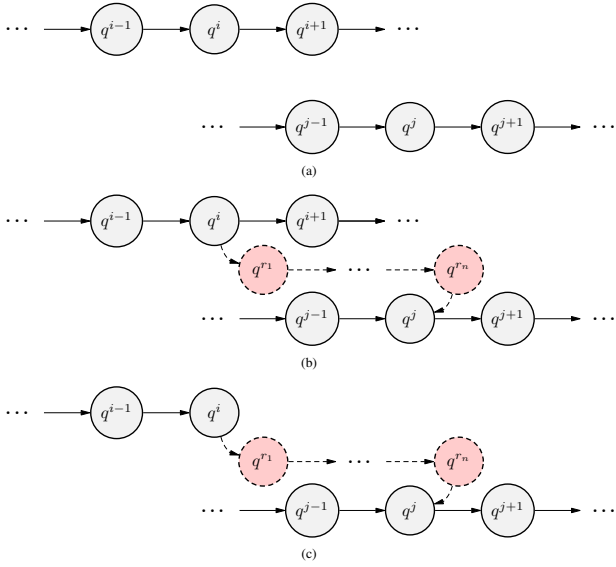

Below definitions describe how failure is acknowledged and reflected in updated PTA used by risk-averse PTA MPC dynamic decision-making algorithm.
% Under Case Auxiliary, $Q^{R}$ is available and path $U_{s1}$ is legal. 
% Finally, under Case Emergency, $Q^{R}$ is available but the path $U_{s1}$ is illegal.
\begin{definition}[Emergency Declaration Signal]
The emergency declaration signal (EDS) is defined as follows:
\begin{equation}
    \textbf{Z}=\{z_{i}|1\leq i\leq |E|, z_{i}\in \{-1,0,1\} \}
\end{equation}
where $z_{i}=-1$, $z_{i}=0$ and $z_{i}=1$ correspond to Case Auxiliary, Case Normal and Case Emergency, respectively.
\end{definition}
Figure \ref{figure-original,aux and emergency paths} depicts two equivalent paths connected with a redundant path and different values of $z_{i}$, where $i$ is the index of the state the redundant path starts from. 
As shown in Figure \ref{figure-original language}, under Case Normal , the redundant path(s) associated with state $q^{i}$ is/are excluded and the system operates by original automaton. 
Figure \ref{figure-auxiliary language} shows the system under Case Auxiliary where the redundant path is added. As per Figure \ref{figure-emergency language}, for Case Emergency the part of the path past $q^{i}$ is removed due to failure and the redundant path is added.

We also introduces a new operator to modify the Original Automaton $A^{O}$ based on the EDS and to return an updated PTA based on possible state failures.
\begin{definition}[Update Operator]
The update operator $\Omega$ acts on the components of the automaton $\mathcal{A}$, in order to reflect the changes made by adding the redundant path(s) $\| \cdot _{r} \|$, to the original one $\| \cdot _{o} \|$, and by trimming illegal path components $\| \cdot _{trim} \|$ that are disconnected from the initial state, i.e. the state(s) which no path exists connecting them to the initial state.
\begin{equation}
\label{eq-update operator}
\begin{split}
    & \Omega(\| \cdot _{o} \|,\| \cdot _{r} \|,\| \cdot _{trim} \|,z_{i},\mathcal{I}_{\boldsymbol{c_{i+1}}})=\\
    &\begin{cases}
    \| \cdot _{o} \| & (z_{i}=0) \cup (\mathcal{I}_{\boldsymbol{c_{i+1}}}=\text{True})\\
    (\| \cdot _{o} \| \cup \| \cdot _{r} \|)- \| \cdot _{trim} \|& \text{otherwise}
    \end{cases}
\end{split}
\end{equation}
where $i$ is the index of the first state of redundant path $\forall z_{i}\in \textbf{Z}, \forall 1\leq i \leq|E|$, and $\mathcal{I}_{c_{i+1}}$ is the $(i+1)^{th}$ element of the invariant operator showing if the state with the path has already been occupied (traversed). Thus, the update operator keeps the original PTA components if the EDS indicates Case Normal, or if the state with redundant path is already traversed. Because the graph of PTA is assumed to be simple~\cite{balta_model_2022}, the failure in states that have already been traversed will not change the PTA.
\end{definition}

\begin{definition}[Active Redundant Path]
The redundant path $Q^{R}=\langle q^{i},q^{r_{1}},\cdots,q^{r_{n}},q^{j}\rangle$ that connects equivalent paths $U_{1}=\langle \cdots ,q^{i-1},q^{i},q^{i+1},\cdots \rangle$ and $U_{2}=\langle \cdots ,q^{j-1},q^{j},q^{j+1},\allowbreak \cdots \rangle$ is defined to be an active redundant path if the two subsets $U_{m1}=\langle \cdots,q^{i-1},q^{i}\rangle$ and $U_{m2}=\langle q^{j},q^{j+1},\cdots \rangle$ are legal paths. Otherwise the redundant path is passive. Note that the redundant path status is time-varying and depends on the updates generated by the update operator based on EDS of paths prior to the path of interest.
\end{definition}
Equation \ref{eq-update operator} is implicitly a function of time, as the Boolean invariant operator $I_{c_{i+1}}$ is changing value over time depending on clock $c_{i+1}$. This relates the formulation in this work to the one developed for PTA MPC without risk-averse feature, with the exception that  the update operator updates the PTA prior to running the PTA MPC algorithm. For further details about the invariant operator and clocks see references \cite{balta_model_2022,kovalenko_cooperative_2022}. 
\par
Assume the original system PTA is given by $\mathcal{A}^{O}=(Q^{O}, C^{O},\Sigma^{O}, E^{O}, I^{O}, R^{O}, P^{O}, q_{0})$. Then each state is assigned with an EDS, with the default value of $\textbf{Z}_{1\times |Q^{O}|}=0$ where the suffix shows array dimension.
\par
Case Normal represents the PTA tuple under normal condition, i.e. $\exists i\in \{1,\cdots,|E|\}: z_{i}=0$, is defined as follows:
\begin{equation}
    \begin{split}
        &\mathcal{A^N}=\Omega(\mathcal{A}^{O},\mathcal{A}_{r},\emptyset,0,\mathcal{I}_{c_{i+1}})\\
    &=\mathcal{A}^{O}=(Q^{O}, C^{O},\Sigma^{O}, E^{O}, I^{O}, R^{O}, P^{O}, q_{0})\\
    \end{split}
\end{equation}
    
Therefore, under Case Normal, i.e. $\forall t,\forall z_{i}\in \textbf{Z}: z_{i}=0$, the equivalent paths of the risk-averse PTA are exactly the same as original PTA.
\par
Case Auxiliary is when the original PTA is still available while redundant path provides an alternative path to promote flexibility in the system. If the PTA is traversed by more than one occupant at a time, the Case Auxiliary provides a detour to avoid state collision, i.e. occurrence of more than one occupant in the same state at the same time. The Case Auxiliary, i.e. $\exists i\in \{1,\cdots,|E|\}: z_{i}=-1$, is defined as follows:
\begin{equation}
\begin{split}
     &\mathcal{A^A}=\Omega(\mathcal{A}^{O},\mathcal{A}_{r},\emptyset,-1,\mathcal{I}_{c_{i+1}})\\
     &=(Q^{A}, C,\Sigma, E^{A}, I, R, P^{A}, q_{0})\\
\end{split}
\end{equation}
where:
\begin{align}
    & Q^{A}=(Q^{O},\langle q^{r^{i}_{1}}, \cdots,q^{r^{i}_{n}}\rangle) \notag\\
    & E^{A}=(E^{O},\langle e^{r^{i}_{1}}, \cdots,e^{r^{i}_{n}}\rangle) \label{eq-Ea}\\
    & P^{A}=(P^{O},\langle p^{r^{i}_{1}}, \cdots,p^{r^{i}_{n}}\rangle) \notag
\end{align}
In this scenario, the redundant path's PTA components are concatenated to the original PTA resulting in a new PTA $\mathcal{A^{A}}$. Note that when Case Auxiliary is declared, the equivalent path past the state with auxiliary path are still \textit{legal}, therefore, $\| \cdot _{trim} \|=\emptyset$.
The rest of the PTA elements of $\mathcal{A^{A}}$ are the same as their corresponding original PTA.
\par
The Case Emergency, $\exists i\in \{1,\cdots,|E|\}: z_{i}=1$, occurs when some states in the original PTA become \textit{illegal}.
In this case, the redundant path provides a new path to avoid MPC failure. Therefore, the update operator \textit{modifies}  the PTA resulting in a new PTA $\mathcal{A^{E}}$. Note that when Case Emergency is declared, $\| \cdot _{trim} \|$ is no longer empty:
\begin{equation}
\begin{split}
    &\mathcal{A^E}=\Omega(\mathcal{A}^{O},\mathcal{A}_{r},\mathcal{A}_{trim},1,\mathcal{I}_{c_{i+1}})\\
    &=(Q^{E}, C,\Sigma, E^{E}, I, R, P^{E}, q_{0})\\
\end{split}
\end{equation}
where $\mathcal{A}_{trim}$ is the part of automaton to be trimmed and the rest of the terms are as follows:
\begin{align}
    & Q^{E}=(Q^{O},\langle q^{r^{i}_{1}}, \cdots,q^{r^{i}_{n}}\rangle)-\langle q^{i+1},q^{i+2},\cdots \rangle \notag\\
    & E^{E}=(E^{O},\langle e^{r^{i}_{1}}, \cdots,e^{r^{i}_{n}}\rangle)-\langle e^{i+1},e^{i+2},\cdots \rangle\\
    & P^{E}=(P^{O},\langle p^{r^{i}_{1}}, \cdots,p^{r^{i}_{n}}\rangle)-\langle p^{i+1},p^{i+2},\cdots \rangle \notag
\end{align}
Similar to the case-auxiliary, the remaining elements of $\mathcal{A^{E}}$ are defined according to the above equations.

\section{Risk-Averse PTA MPC}
\label{sec-PTA MPC}
%In PTA, the time elapses independent of any event.
The control objective for the PTA-MPC algorithm is to find a path that minimizes the sum of all temporal cost. For PTA-MPC, the time elapsed at each state execution and state transition is stored as a clock variable. The temporal cost is calculated based on these clock variables. As a result, a finite path, i.e. a finite ordered sequence of states, is associated with a deterministic temporal cost which enables formulating a constrained optimization problem. 
% The control objective is to find a path such that the sum of all cost metrics for states comprising the path is minimized.
\par
Figure \ref{figure-MPC_block diagram} depicts the risk-averse PTA MPC block diagram. 
At each iteration the update operator generates the updated PTA using the original PTA, redundant PTA, and EDS. The current state and remaining desired state set are retrieved from memory and used by the update operator and the risk-averse PTA MPC algorithm, respectively. The risk-averse PTA MPC also receives  user defined risk-factors for each state. These risk-factors are designated as $h_{i}$   $\forall i\in \{1,\cdots,|Q^{T}|\}$ and determine the weight of failure risk at each state. Thus, the risk factors provide a quantitative measure to compare paths in terms of risk.
\par
The goal of the multi-objective optimization problem is to minimize the cost of the path while minimizing the risk, which can be represented as follows:
\begin{equation}
\label{eq-cost & risk optimization-general}
%\alpha^{\ast}= \underset{\alpha \in \mathcal{L}(\mathcal{A})}{\arg\min } \left(V(\alpha) =\sum_{i=1}^{|\alpha|}P_{i}+\sum_{i=1}^{|\alpha|}R_{i}\right),  
\alpha^{\ast}= \underset{\alpha \in \mathcal{L}(\mathcal{A})}{\arg\min } \left(V(\alpha) =\sum_{i=1}^{|\alpha|}P_{i},\sum_{i=1}^{|\alpha|}R_{i}\right),  
\end{equation}
where $\alpha$ is the path, i.e. the sequence of states, and $R_{i}$ is the \textit{Risk Measure} of the $i^{th}$ state showing the associated risk weight corresponding to each state, and $\mathcal{L}(\mathcal{A})$ is the set of all feasible paths in PTA $\mathcal{A}$. The risk measure is comprised of two parts; user-defined weights and PTA-inherited weight. The user-defined weights indicate the risk chance as determined by the user whereas the PTA-inherited weight is based on the number of redundant paths available.
\par
While Equation \ref{eq-cost & risk optimization-general} can be solved as a multi-objective optimization problem, it can be reduced to a single-objective optimization problem if the risk measures can be described with respect to the state cost, i.e. $R_{i}=u_{i}P_{i}$. In this manner, the risk measure can be expressed relative to the cost measure as follows: 
\begin{align}
& \alpha^{\ast}= \underset{\alpha \in \mathcal{L}(\mathcal{A})} {\arg\min } \left( V(\alpha) = \sum_{i=1}^{|\alpha|}P_{i}+\sum_{i=1}^{|\alpha|}u_{i}P_{i}\right). \label{eq-cost function}
% & subject\hspace{0.2 cm}to: \alpha \in \mathcal{L}(\mathcal{A}) \notag
\end{align}
The variable $u_{i}$ captures the \textit{uncertainty ratio} associated with $i^{th}$ state of the path which is defined as follows:
\begin{equation}
\label{eq-uncertainty ratio}
    u_{i}=\frac{h_{i}}{x_{i}}
\end{equation}
where $h_{i}$ is the risk factor assigned by the user to each state and $x_{i}$ is \textit{Out-degree Centrality Measure} defined as being the number of outgoing edges of a given state \cite{newman_networks_2018}. 
% Since the last state of each path is the last desired state in the desired state set $\Sigma$ and thus, exists in all possible solution and has a zero out-degree centrality, it is excluded from the risk sum.
We exclude the last state of each path from the risk sum since it is assumed to have centrality zero and it belongs to every possible path.
Assuming that $\alpha$ is a path of length N, i.e. $\alpha=\langle \alpha_{i_{1}},\cdots, \alpha_{i_{N}} \rangle$ and $\Sigma=\langle \sigma_{1},\cdots,\sigma_{N_{d}} \rangle$ with $N_{d}\leq N$, where $N_{d}$ is the number of desired states, we combine Equation~\ref{eq-cost function} and  Equation~\ref{eq-uncertainty ratio}  to obtain:
\begin{align}
\label{eq-cost function- modified}
&\alpha^{\ast}= \underset{\alpha}{\arg\min} \left( V(\alpha)= \sum_{i=1}^{|\alpha|-1}(1+\frac{h_{i}}{x_{i}})P_{i}\right)\\
&\alpha \in \mathcal{L}(\mathcal{A}) \label{eq-validity}\\
&\forall m<n \in \{1,\cdots,N_{d}\},\forall \sigma_{m},\sigma_{n} \in \Sigma \notag\\
&\exists \alpha_{ik},\alpha_{il} \in \alpha: \sigma_{m}=\alpha_{ik},\sigma_{m}=\alpha_{il},k<l\hspace{0.5cm}\label{eq-desired value}
\end{align}

\begin{figure}[t]
\begin{center}
    \begin{adjustbox}{minipage=\textwidth,scale=0.7}
    \begin{tikzpicture}[>={Stealth[inset=0pt,length=8pt,angle'=28,round]},scale=1,font=\normalsize]
      % coordinates
        \coordinate (LLT) at (-1.65,-3);
        \coordinate (LLMPC) at (1,4);
        \coordinate (LLRPTA) at (4,-0.5);
        \coordinate (LLTPTA) at (3.9,-0.6);
        \coordinate (LLZ) at (-3,-0.5);
        \coordinate (LLLE) at (6,-1.75);
        \coordinate (LLOPTA) at (4,1);
        \coordinate (LLLN) at (6,0.75);
        \coordinate (LLUO) at (1,0);
        \coordinate (LLM) at (1,2);
        \coordinate (LLFP) at (-3,4.5);
      % nodes
        \node[draw, minimum width=2.45cm, minimum height=2.7cm, anchor=south west, text width=2cm, font=\normalsize, align=center,label=0:$ \mathcal{A}^{T}$, densely dashed] (TPTA) at (LLTPTA)
        {};
        
        \node[draw, minimum width=2cm, minimum height=1.5cm, anchor=south west, text width=2cm, font=\normalsize, align=center] (FP) at (LLFP)
        {User Defined\\Risk-Factors};
        
        \node[draw, minimum width=2cm, minimum height=1cm, anchor=south west, text width=2cm, font=\normalsize, align=center] (EDS) at (LLZ)
        {EDS (\textbf{Z})};
      
        \node[draw, minimum width=2cm, minimum height=1.5cm, anchor=south west, text width=2cm, font=\normalsize, align=center] (MPC) at (LLMPC)
        {Risk-Averse\\PTA-MPC};
        
        \node[draw, minimum width=2cm, minimum height=1cm, anchor=south west, text width=2cm, font=\normalsize, align=center] (LT) at (LLOPTA) {Original\\PTA  ($\mathcal{A}^{O}$)};
        
        \node[draw, minimum width=2cm, minimum height=1cm, anchor=south west, text width=2cm, font=\normalsize, align=center] (Z) at (LLRPTA) {Redundant\\PTA ($\mathcal{A}^{R}$)};
        
        \node[draw, minimum width=2cm, minimum height=1.5cm, anchor=south west, text width=2cm, font=\normalsize, align=center] (UO) at (LLUO) {Update\\Operator ($\Omega$)};
        
        \node[draw, minimum width=2cm, minimum height=1cm, anchor=south west, text width=2cm, font=\normalsize, align=center] (M) at (LLM) {Memory};
        
        %edges
        % MPC
        \draw[->] (MPC.0) -- ($(MPC.0)+(2,0)$) node[anchor=center,above] {$\bar{\alpha}(t+1)$};
        \path[draw,-] ($(MPC.0)+(0,-0.5)$) -- ($(MPC.0)+(1,-0.5)$);
        \draw[->] ($(MPC.0)+(1,-0.5)$) |- node[left,pos=0.3]{$D(t+1)$} ($(M.0)+(0,0.2)$);
        \draw[->] ($(MPC.0)+(1.5,0)$) |-  ($(M.0)+(0,-0.2)$);
        \draw[->] ($(M.270)$) -- (UO.90)node[right,pos=0.5] {$\bar{\alpha}(t)$};
        
        \path[draw,-] ($(UO.180)+(0,0.2)$) -- ($(UO.180)+(-1.4,0.2)$);
        \draw[->] ($(UO.180)+(-1.4,0.2)$) |- node[left,pos=0.25]{\rotatebox{90}{Updated PTA}} (MPC.180);
        
        % Memory
        \path[draw,-] (M.180) -- ($(M.180)+(-0.9,0)$);
        \draw[->] ($(M.180)+(-0.9,0)$) |- node[right,pos=0.25]{$D(t)$} ($(MPC.180)+(0,-0.5)$);
        
        % Risk Factor
        
        \draw[->] (FP.0) -- ($(MPC.180)+(0,0.5)$);
        
        %Languages
        \path[draw,-] (LT.180) -- ($(LT.180)+(-0.3,0)$);
        \path[draw,-] ($(LT.180)+(-0.3,0)$) -- ($(LT.180)+(-0.3,-0.55)$);
        \draw[->] ($(LT.180)+(-0.3,-0.55)$) -- ($(UO.0)+(0,0.2)$);
        
        \path[draw,-] (Z.180) -- ($(Z.180)+(-0.3,0)$);
        \path[draw,-] ($(Z.180)+(-0.3,0)$) -- ($(Z.180)+(-0.3,0.55)$);
        \draw[->] ($(Z.180)+(-0.3,0.55)$) -- ($(UO.0)+(0,-0.2)$);
        
        \path[draw,-] (EDS.0) -- ($(EDS.0)+(0.3,0)$);
        \path[draw,-] ($(EDS.0)+(0.3,0)$) -- ($(EDS.0)+(0.3,0.55)$);
        \draw[->] ($(EDS.0)+(0.3,0.55)$) -- ($(UO.180)+(0,-0.2)$);
        
    \end{tikzpicture}    
    \end{adjustbox}
    \end{center}
    \caption{Block diagram of risk-averse MPC with update operator}
    \label{figure-MPC_block diagram}
\end{figure}
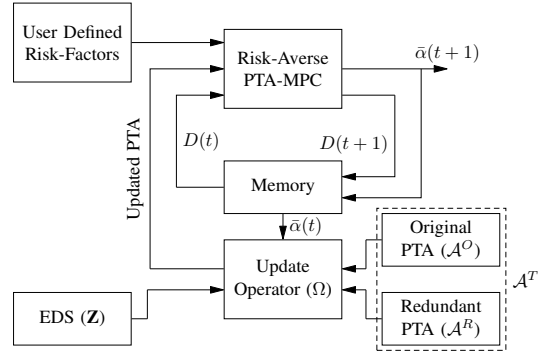
\begin{algorithm}[t]
\caption{Risk-Averse PTA-MPC Optimization Algorithm}
\label{algorithm-MPC}
\begin{algorithmic}[1]
\Require{$\mathcal{A}=(Q, C,\Sigma, E, I, R, P, q_{0}),D(t),h_{i}\hspace{0.2cm}\forall\hspace{0.1cm}1\leq i\leq|E|$}
\Ensure{$\alpha(t+1),D(t+1)$}
\Initialize{
$\mathcal{A}_{updated} \gets \mathcal{A}$\\
$D(t+1) \gets \Sigma$,\hspace{0.1cm}$\alpha(t+1) \gets \emptyset$\\
$\bar{\alpha}(t) \gets q_{0}$,\hspace{0.1cm}$x \gets \emptyset$\\
}
\While{$D(t+1)\neq \emptyset$}
\State (i) Check {\fontfamily{qcr}\selectfont SAT} for Eq. \ref{eq-cost function- modified}
\If{(i) is {\fontfamily{qcr}\selectfont SAT}}
\State Determine risk-averse PTA-MPC optimization solution $\bar{\alpha}(t+1)$ based on Eq. \ref{eq-cost function- modified} using $\mathcal{A}_{updated}$ and $h_{i}$s
\State $\alpha(t+1) \gets \langle \alpha(t),\bar{\alpha}[1](t+1) \rangle$
\State \Comment{$\bar{\alpha}[1](t+1)$ is the first element of $\bar{\alpha}(t+1)$}
\Else
\State return {\fontfamily{qcr}\selectfont UNSAT}
\State stop
\EndIf
\State $\bar{\alpha}(t) \gets \bar{\alpha}[1](t+1)$
\If{$\bar{\alpha}[1](t+1) \in \Sigma$}
\State $D(t+1) \gets \Sigma-{\bar{\alpha}[1](t+1)} $
\EndIf
\State $D(t) \gets D(t+1)$
\State return $\alpha(t+1),D(t+1)$
\EndWhile
\State return {\fontfamily{qcr}\selectfont SAT}
\end{algorithmic}
\end{algorithm}
Equation \ref{eq-cost function- modified} describes the risk measure in term of cost. Equation \ref{eq-validity} ensures that the solution is among feasible paths of $\mathcal{A}$. Equation \ref{eq-desired value} ensures that the solution includes all desired states according to $\Sigma$.
Regardless of the value of EDS, at all times the PTA and its constraints are the same as in \cite{balta_model_2022}. Thus, the optimization problem described by Equation \ref{eq-cost function- modified} is reduced to first-order logic, which can be solved using the $Z_{3}$ theorem prover \cite{bjorner_z_nodate}.
Algorithm \ref{algorithm-MPC} summarizes the internal risk-averse PTA-MPC optimization mechanism. First, the EDS is read and based on the current state, the active redundant paths are identified. Then, the PTA is modified by the update operator. The satisfiability solver is used to check if step (i) is satisfiable ({\fontfamily{qcr}\selectfont SAT}). If so, the optimization solution is determined based on the updated PTA and user defined risk factors $h_{i}$, and the first element of solution array is appended to the actual solution. Otherwise, the optimization is unsatisfiable ({\fontfamily{qcr}\selectfont UNSAT}). If the first element of the solution belongs to the desired set of states, then it is removed from the remaining set of desired states.
The iteration continues until either the remaining set is empty or optimization fails.

%\subsection{Multi-Product Optimization}
%So far, we assumed that the optimization problem involves only a single product, which implies that all states and edges are assumed to be clear except for the state or edge being occupied by the product itself.
%Now, assume that products $R_{1}$ and $R_{2}$ are taking the path simultaneously, being the leading and trailing in time, respectively. Also, assume that the EDS is declaring an state other that normal and that product $R_{1}$ has taken the emergency path already, and that $R_{2}$ starts taking the path with $\Delta t$ ticks of delay with respect to $R_{1}$, i.e. $t_{1}=t_{2}+\Delta t$. Below, we are studying the conditions under which state coincidence is avoided:
%\begin{equation}
%    q_{R_{1}}(t_{1}) \neq q_{R_{2}}(t_{2})
%\end{equation}
%We may use equations\ref{eq-total time} and \ref{eq-remaining time} to rewrite the above equation:
%\begin{equation}
%\label{eq-delay condition}
%    D^{q(R_{1})}+\tau_{r_{R_{1}}}\leq  \Delta t + D^{q(R_{2})}
%\end{equation}
%where $D^{q(R_{1})}$ and $D^{q(R_{2})}$ are the current state delay of product $R_{1}$ and $R_{2}$, respectively.
%If equation\ref{eq-delay condition} is not satisfied and the state of emergency is declared, i.e. $z=1$, then we need to introduce an \textit{artificial delay} as follows:
%\begin{equation}
%D^{artf}= \Delta t + D^{q(R_{2})}- (D^{q(R_{1})}+\tau_{r_{R_{1}}})     
%\end{equation}
%which will be added to the state delay of the first state in the emergency path in order to prevent collision. Under the state of auxiliary, the introduction of $D^{artf}$ is optional.

PTA MPC without risk-averse feature which we refer to as \textit{PTA MPC}, runs multiple simulation after each event to find optimal path \cite{balta_model_2022}. If the PTA MPC optimization problem does not have a solution due to state failure, the PTA MPC returns \textit{unsatisfiable} ({\fontfamily{qcr}\selectfont
UNSAT}). Otherwise, it returns satisfiable ({\fontfamily{qcr}\selectfont
SAT}), and outputs the optimal path $\bar{\alpha}$, and remaining desired state set $D(t+1)$. The actual path that is traversed after each iteration comprises of the first state of $\bar{\alpha}$, i.e. $\alpha(t+1) \gets \langle \alpha(t)+\bar{\alpha}[1](t+1)\rangle$. This, constitutes the basic of risk-averse PTA MPC as well.
\par
Similar to PTA-MPC, risk-averse PTA-MPC runs the optimization algorithm at many iterations and returns the optimal solution at each iteration based on the PTA updates. Here, the actual solution is labeled as $\alpha(t)$ which appends the first element of optimal solution $\bar{\alpha}(t)$ found by the risk-averse PTA-MPC algorithm at each iteration. 
The optimization continues until the desired state set is exhausted. 
% As time goes by, the goal states surpassed are removed from the desired state set and the optimization continues until the desired state set is exhausted. 
At each iteration, the remaining set of desired states is labelled as $D(t)$. 
If $\forall t: D(t)\subseteq\Sigma$ represent the remaining desired state set at $t_{i}\in t$, then if $D(t)\nsubseteq Q^{E}$, the risk-averse PTA-MPC fails to reach the goal state, i.e. $\bar{\alpha}(t)$ the optimal state array determined by risk-averse PTA-MPC returns {\fontfamily{qcr}\selectfont
UNSAT}. 
Because Equation~\ref{eq-update operator} depends on the optimal state array, the $\bar{\alpha}(t)$ and $D(t)$ are stored in the memory for the next iteration.
\par
Note that for PTA with multiple redundant paths, i.e. $|\textbf{Z}|>1, \exists z_{i},z_{j}\in \textbf{Z}: z_{i},z_{j} \neq 0$, the update operator acts on active redundant paths only. Therefore, if $q^{c}\in Q$ is the current state, the PTA is not changed by redundant paths starting at previous states $q^{i}\in\{q^{1},\cdots, q^{c-1}\}$ regardless of their corresponding EDS value and status (active or not).
To determine if the redundant path $Q^{R}_{i}$ is active, all redundant paths with starting state closer to the initial state are updated using the update operator. That is, $\forall q^{j}, \alpha_{j}=\langle q^{0}, \cdots, q^{j}\rangle , \alpha_{i}=\langle q^{0}, \cdots, q^{i}\rangle: |\alpha_{j}|\leq|\alpha_{i}|$. For assumptions regarding PTA structures controllable by MPC and MPC optimization algorithm see \cite{balta_model_2022}.
\par
If the optimization has no feasible solution, then the algorithm returns \textit{unsatisfiable} ({\fontfamily{qcr}\selectfont UNSAT}). Otherwise, the optimization returns $\alpha(t+1)$ and updates $D(t+1)$ and returns \textit{satisfiable} ({\fontfamily{qcr}\selectfont SAT}).

\section{Case Study}
\label{sec-case study}
Recently, PTAs have been proposed  for product scheduling and product-agent coordination in manufacturing where machine physical locations represent states, the required process time is the cost metric, and product transport/storage represent state transition \cite{kovalenko_model-based_2019,kovalenko_priced_2020,kovalenko_cooperative_2022}. The redundant paths represent manufacturing line alternation means in order to isolate a certain equipment for regular maintenance or replacement. In addition, the redundant paths can be used to prevent product coincidence in one manufacturing equipment when the manufacturing line is used by two products, simultaneously. It is important to ensure  flexibility for product scheduling since customers can impose hard deadlines that need to be met by the manufacturer \cite{ocker_framework_2019,kovalenko_cooperative_2022}.
Thus, the proposed risk-averse PTA MPC framework should be used to ensure that products meet customer deadlines even in the event of machine failures in the manufacturing system.

\begin{figure}[t!]
    \begin{center}
    \begin{adjustbox}{minipage=\textwidth,scale=0.8}

         \subfigure[]{\label{fig-PTA MPC}\begin{tikzpicture}[node distance = 2cm, on grid,>={Stealth[inset=0pt,length=6pt,angle'=28,round]},fill fraction/.style n args={2}{path picture={
            \fill[#1] (path picture bounding box.south west) rectangle
            ($(path picture bounding box.north west)!#2!(path picture bounding box.north east)$);}}
    ]
  %  \tikzstyle{every state}=[opacity=1]

\begin{customlegend}[legend style={at={(4,4)},anchor=north},legend columns=-1,legend style={column sep=1ex},legend entries={risk-averse PTA-MPC,PTA-MPC,Failed State}]
    \addlegendimage{black,only marks, mark=*, fill=green!30,style={scale=4}}
    \addlegendimage{black,only marks, mark=*, fill=magenta!50,style={scale=4}}
    \addlegendimage{black,only marks, mark=*, pattern=north west lines, pattern color=black,style={scale=4}}
   \end{customlegend};
    
    \node[state, initial, opacity=1, fill=magenta!50 , fill fraction={green!30}{0.5}] (q1) {$q^1$};
    \node[state] at (1,-2.5) (q2) {$q^2$};
    \node[state, right of=q2] (q3) {$q^3$};
    \node[state, right of=q3] (q4) {$q^4$};
    \node[state, opacity=1, right of=q4,pattern=north west lines, pattern color=black] (q5) {$q^5$};
    \node[state, opacity=1, fill=magenta!50 , fill fraction={green!30}{0.5}] at (8,0) (q6) {$q^6$};

    \node[state, opacity=1, fill=magenta!50] at (2,0) (q7) {$q^7$};
    \node[state, right of=q7, opacity=1, fill=magenta!50] (q8) {$q^8$};
    \node[state, opacity=1,fill=magenta!50] at (6,0) (q9) {$q^9$};
    
    \node[state, opacity=1, fill=green!30] at (1,2.5) (q10) {$q^{10}$};
    \node[state, right of=q10, opacity=1, fill=green!30] (q11) {$q^{11}$};
    \node[state, right of=q11, opacity=1, fill=green!30] (q12) {$q^{12}$};
    \node[state, right of=q12, opacity=1, fill=green!30] (q13) {$q^{13}$};
    
    %\node[state,fill=red!20, densely dashed] at (2,-1.25) (q14) {$q^{14}$};
    \node[state,fill=red!20, densely dashed] at (4,-1.25) (q15) {$q^{15}$};
    %\node[state, fill=red!20, densely dashed] at (2,1.25) (q16) {$q^{16}$};
    %\node[state, fill=red!20, densely dashed] at (6,1.25) (q17) {$q^{17}$};
\centering
 \draw 
    
   (q1) edge[opacity=1,bend right] node [below left]{$e^1$} (q2)
   (q2) edge[opacity=1] node [below] {$e^2$} (q3)
   (q3) edge[opacity=1] node [below]{$e^3$} (q4)
%   (q4) edge node [below]{$e^4$} (q5)
%   (q5) edge[bend right] node [below right] {$e^5$} (q6)
   
   (q1) edge node [above]{$e^6$} (q7)
   (q7) edge node [above]{$e^7$} (q8)
   (q8) edge node [above]{$e^8$} (q9)
   (q9) edge node [above]{$e^9$} (q6)
   
   (q1) edge[bend left] node [below left]{$e^{10}$} (q10)
   (q10) edge node [above]{$e^{11}$} (q11)
   (q11) edge node [above]{$e^{12}$} (q12)
   (q12) edge node [above]{$e^{13}$} (q13)
   (q13) edge[bend left] node [above right]{$e^{14}$} (q6)
   
   %(q2) edge[bend right, densely dashed,opacity=1] node [above left]{$e^{15}$} (q14)
   %(q14) edge[bend right, densely dashed,opacity=1] node [ right]{$e^{16}$} (q7)
   (q4) edge[bend right, densely dashed,opacity=1] node [ right]{$e^{17}$} (q15)
   (q15) edge[bend right, densely dashed,opacity=1] node [ right]{$e^{18}$} (q9);
   %(q12) edge[bend left=10, densely dashed,opacity=1] node [ above]{$e^{19}$} (q16)
   %(q16) edge[bend left=10, densely dashed,opacity=1] node [ above right]{$e^{20}$} (q9)
   %(q12) edge[bend left=10, densely dashed,opacity=1] node [ above right]{$e^{21}$} (q17)
   %(q17) edge[bend left, densely dashed,opacity=1] node [ above right]{$e^{22}$} (q9)
    \end{tikzpicture}}

        \subfigure[]{\label{fig-risk-averse}\begin{tikzpicture}[node distance = 2cm, on grid,>={Stealth[inset=0pt,length=6pt,angle'=28,round]},fill fraction/.style n args={2}{path picture={
            \fill[#1] (path picture bounding box.south west) rectangle
            ($(path picture bounding box.north west)!#2!(path picture bounding box.north east)$);}}
    ]
    %\tikzstyle{every state}=[opacity=1]
    
    \node[state, initial, opacity=1, fill=magenta!50 , fill fraction={green!30}{0.5}] (q1) {$q^1$};
    \node[state, opacity=1, fill=green!30] at (1,-2.5) (q2) {$q^2$};
    \node[state, right of=q2, opacity=1, fill=green!30] (q3) {$q^3$};
    \node[state, right of=q3, opacity=1, fill=green!30] (q4) {$q^4$};
    \node[state, opacity=1, right of=q4,pattern=north west lines, pattern color=black] (q5) {$q^5$};
    \node[state, opacity=1, fill=green!30] at (8,0) (q6) {$q^6$};

    \node[state, opacity=1, fill=magenta!50] at (2,0) (q7) {$q^7$};
    \node[state, right of=q7,pattern=north west lines, pattern color=black, opacity=1] (q8) {$q^8$};
    \node[state, opacity=1, fill=green!30] at (6,0) (q9) {$q^9$};
    
    \node[state] at (1,2.5) (q10) {$q^{10}$};
    \node[state, right of=q10,pattern=north west lines, pattern color=black, opacity=1] (q11) {$q^{11}$};
    \node[state, right of=q11] (q12) {$q^{12}$};
    \node[state, right of=q12] (q13) {$q^{13}$};
    
    %\node[state,fill=red!20, densely dashed] at (2,-1.25) (q14) {$q^{14}$};
    \node[state, opacity=1, fill=green!30, densely dashed] at (4,-1.25) (q15) {$q^{15}$};
    %\node[state,fill=red!20, densely dashed] at (2,1.25) (q16) {$q^{16}$};
    %\node[state,fill=red!20, densely dashed] at (6,1.25) (q17) {$q^{17}$};
\centering
 \draw 
    
   (q1) edge[bend right] node [below left]{$e^1$} (q2)
   (q2) edge node [below] {$e^2$} (q3)
   (q3) edge node [below]{$e^3$} (q4)
%   (q4) edge node [below]{$e^4$} (q5)
%   (q5) edge[bend right] node [below right] {$e^5$} (q6)
   
   (q1) edge node [above]{$e^6$} (q7)
%  (q7) edge node [above]{$e^7$} (q8)
%  (q8) edge node [above]{$e^8$} (q9)
   (q9) edge node [above]{$e^9$} (q6)
   
   (q1) edge[bend left,opacity=1] node [below left]{$e^{10}$} (q10)
%   (q10) edge node [above]{$e^{11}$} (q11)
%   (q11) edge node [above]{$e^{12}$} (q12)
   (q12) edge[opacity=1] node [above]{$e^{13}$} (q13)
   (q13) edge[bend left,opacity=1] node [above right]{$e^{14}$} (q6)
   
   %(q2) edge[bend right, densely dashed,opacity=1] node [above left]{$e^{15}$} (q14)
   %(q14) edge[bend right, densely dashed,opacity=1] node [ right]{$e^{16}$} (q7)
   (q4) edge[bend right, densely dashed] node [ right]{$e^{17}$} (q15)
   (q15) edge[bend right, densely dashed] node [ right]{$e^{18}$} (q9);
   %(q12) edge[bend left=10, densely dashed,opacity=1] node [ above]{$e^{19}$} (q16)
   %(q16) edge[bend left=10, densely dashed,opacity=1] node [ above right]{$e^{20}$} (q9)
   %(q12) edge[bend left=10, densely dashed,opacity=1] node [ above right]{$e^{21}$} (q17)
   %(q17) edge[bend left, densely dashed,opacity=1] node [ above right]{$e^{22}$} (q9)
    \end{tikzpicture}}

        \subfigure[]{\label{fig-pareto risk-averse}\begin{tikzpicture}[node distance = 2cm, on grid,>={Stealth[inset=0pt,length=6pt,angle'=28,round]},fill fraction/.style n args={2}{path picture={
            \fill[#1] (path picture bounding box.south west) rectangle
            ($(path picture bounding box.north west)!#2!(path picture bounding box.north east)$);}}
    ]
   % \tikzstyle{every state}=[opacity=1]

    \node[state, initial, opacity=1, fill=magenta!50 , fill fraction={green!30}{0.5}] (q1) {$q^1$};
    \node[state] at (1,-2.5) (q2) {$q^2$};
    \node[state, right of=q2] (q3) {$q^3$};
    \node[state, right of=q3] (q4) {$q^4$};
    \node[state, opacity=1, right of=q4,pattern=north west lines, pattern color=black] (q5) {$q^5$};
    \node[state, opacity=1, fill=green!30] at (8,0) (q6) {$q^6$};

    \node[state, opacity=1, fill=magenta!50] at (2,0) (q7) {$q^7$};
    \node[state, right of=q7,pattern=north west lines, pattern color=black, opacity=1] (q8) {$q^8$};
    \node[state, opacity=1, fill=green!30] at (6,0) (q9) {$q^9$};
    
    \node[state, opacity=1, fill=green!30] at (1,2.5) (q10) {$q^{10}$};
    \node[state, right of=q10, opacity=1, fill=green!30] (q11) {$q^{11}$};
    \node[state, right of=q11, opacity=1, fill=green!30] (q12) {$q^{12}$};
    \node[state, right of=q12,pattern=north west lines, pattern color=black, opacity=1] (q13) {$q^{13}$};
    
    %\node[state,fill=red!20, densely dashed] at (2,-1.25) (q14) {$q^{14}$};
    \node[state,fill=red!20, densely dashed] at (4,-1.25) (q15) {$q^{15}$};
    \node[state, opacity=1, fill=green!30, densely dashed] at (2,1.25) (q16) {$q^{16}$};
    \node[state, opacity=1, fill=red!20, densely dashed] at (6,1.25) (q17) {$q^{17}$};
\centering
 \draw 
    
   (q1) edge[opacity=1,bend right] node [below left]{$e^1$} (q2)
   (q2) edge[opacity=1] node [below] {$e^2$} (q3)
   (q3) edge[opacity=1] node [below]{$e^3$} (q4)
%   (q4) edge node [below]{$e^4$} (q5)
%   (q5) edge[bend right] node [below right] {$e^5$} (q6)
   
   (q1) edge node [above]{$e^6$} (q7)
%  (q7) edge node [above]{$e^7$} (q8)
%  (q8) edge node [above]{$e^8$} (q9)
   (q9) edge node [above]{$e^9$} (q6)
   
   (q1) edge[bend left] node [below left]{$e^{10}$} (q10)
   (q10) edge node [above]{$e^{11}$} (q11)
   (q11) edge node [above]{$e^{12}$} (q12)
%   (q12) edge[opacity=1] node [above]{$e^{13}$} (q13)
%   (q13) edge[bend left,opacity=1] node [above right]{$e^{14}$} (q6)
   
   %(q2) edge[bend right, densely dashed,opacity=1] node [above left]{$e^{15}$} (q14)
   %(q14) edge[bend right, densely dashed,opacity=1] node [ right]{$e^{16}$} (q7)
   (q4) edge[bend right, densely dashed,opacity=1] node [ right]{$e^{17}$} (q15)
   (q15) edge[bend right, densely dashed,opacity=1] node [ right]{$e^{18}$} (q9)
   (q12) edge[bend left=10, densely dashed] node [ above]{$e^{19}$} (q16)
   (q16) edge[bend left=10, densely dashed] node [ above right]{$e^{20}$} (q9)
   (q12) edge[bend left=10, densely dashed] node [ above right]{$e^{21}$} (q17)
   (q17) edge[bend left, densely dashed] node [ above right]{$e^{22}$} (q9);
    \end{tikzpicture}}

    \end{adjustbox}
    \end{center}
    \caption{The optimal path for risk-averse PTA MPC  (green) versus PTA MPC (magenta) in presence of state failure (hashed) for:
    \subref{fig-PTA MPC} Scenario $1$: Initial failure at $q^{5}$ only,
    \subref{fig-risk-averse} Scenario $2$: Initial failure at $q^{11}$ and later failure at $q^{5}$ and $q^{8}$ and, \subref{fig-pareto risk-averse} Scenario $3$: Initial failure at $q^{5}$ and later failure at $q^{8}$ and $q^{13}$.}
    \label{figure-sys comparison}
\end{figure}
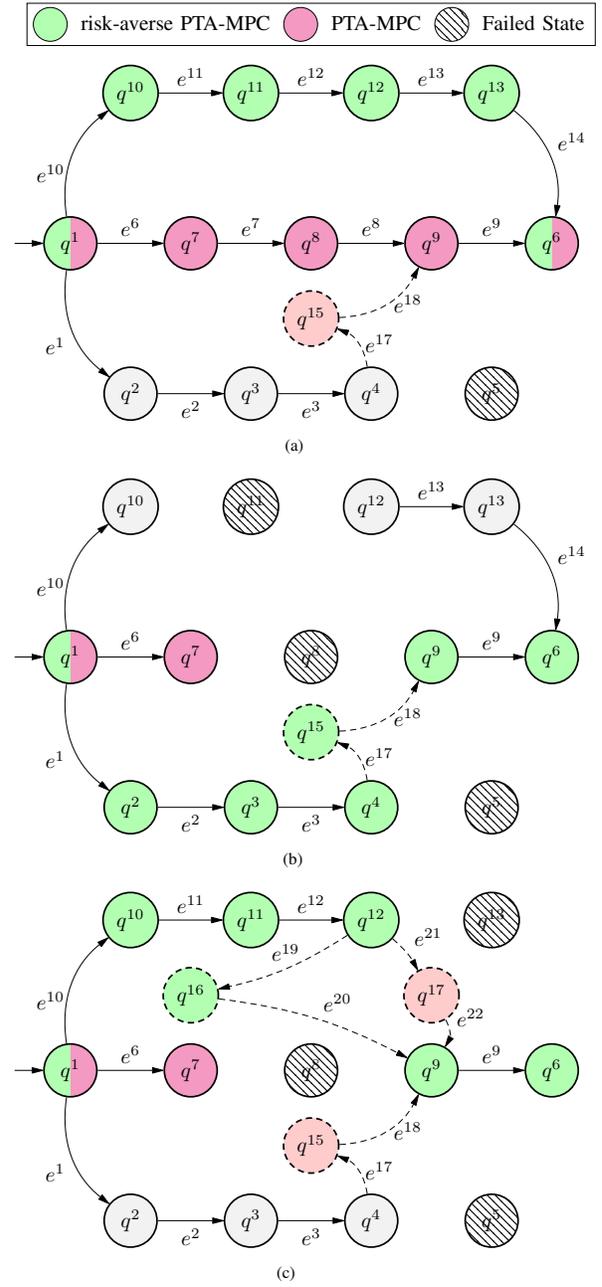

Figure~\ref{figure-sample PTA} depicts a typical layout of manufacturing system in which Original Automaton $\mathcal{A}^{O}$ and Redundant Automaton $A^{R}$ are marked by solid lines and dashed lines, respectively. Table \ref{tab:list of locations for systems} summarizes the physical locations, their associated costs and corresponding user-defined risk factors. For this example, $\Sigma=\{\sigma_{1}^{d}\}=\{q^{6}\}$ and we compare the results of risk-averse PTA MPC to that of PTA MPC for three scenarios:
\begin{itemize}[leftmargin=*]
    \item Scenario 1:
    $z_{5}(t\geq0)=1,$
    \item Scenario 2: $z_{11}(t\geq0)=z_{5}(t\geq t_{1})=z_{8}(t\geq t_{2})=1,t_{1}>t_{2},$
    \item Scenario 3:
    $z_{5}(t\geq0)=z_{13}(t\geq t_{1})=z_{8}(t\geq t_{2})=1,t_{1}>t_{2},$
    
\end{itemize}
Figure~\ref{figure-sys comparison} compares the proposed risk-averse PTA MPC performance with that of previously developed PTA MPC~\cite{balta_model_2022}.
\par
In Scenario 1, failure occurs initially at $q^{5}$ which enables both controller to reach the desired state without a need to enable redundant paths. In this case, the optimal path and the objective function value for PTA MPC and risk-averse PTA MPC are $\alpha_{PM}^{\ast}=\langle q^{1},q^{7},q^{8},q^{9},q^{6} \rangle$ for $V^{\ast}=7.33$ and $\alpha_{RAPM}^{\ast}=\langle q^{1},q^{10},q^{11},q^{12},q^{13},q^{6} \rangle$ for $V^{\ast}=8.66$, respectively. Therefore, in Scenario 1 the PTA MPC has a lower optimal objective function value suggesting that in case of no-failure, despite choosing a safer path, risk-averse PTA MPC might have a higher cost objective function value.

\begin{figure}[t]
    \begin{center}
        \begin{adjustbox}{minipage=\textwidth,scale=0.8}
    \begin{tikzpicture}[node distance = 2cm, on grid,>={Stealth[inset=0pt,length=6pt,angle'=28,round]}]
    
    \node[state, initial] (q1) {$q^1$};
    \node[state] at (1,-2.5) (q2) {$q^2$};
    \node[state, right of=q2] (q3) {$q^3$};
    \node[state, right of=q3] (q4) {$q^4$};
    \node[state, right of=q4] (q5) {$q^5$};
    \node[state] at (8,0) (q6) {$q^6$};

    \node[state] at (2,0) (q7) {$q^7$};
    \node[state, right of=q7] (q8) {$q^8$};
    \node[state] at (6,0) (q9) {$q^9$};
    
    \node[state] at (1,2.5) (q10) {$q^{10}$};
    \node[state, right of=q10] (q11) {$q^{11}$};
    \node[state, right of=q11] (q12) {$q^{12}$};
    \node[state, right of=q12] (q13) {$q^{13}$};
    
    \node[state,fill=red!20, densely dashed] at (2,-1.25) (q14) {$q^{14}$};
    \node[state,fill=red!20, densely dashed] at (4,-1.25) (q15) {$q^{15}$};
    \node[state,fill=red!20, densely dashed] at (2,1.25) (q16) {$q^{16}$};
    \node[state,fill=red!20, densely dashed] at (6,1.25) (q17) {$q^{17}$};

 \draw 
    
   (q1) edge[bend right] node [below left]{$e^1$} (q2)
   (q2) edge node [below] {$e^2$} (q3)
   (q3) edge node [below]{$e^3$} (q4)
   (q4) edge node [below]{$e^4$} (q5)
   (q5) edge[bend right] node [below right] {$e^5$} (q6)
   
   (q1) edge node [above]{$e^6$} (q7)
   (q7) edge node [above]{$e^7$} (q8)
   (q8) edge node [above]{$e^8$} (q9)
   (q9) edge node [above]{$e^9$} (q6)
   
   (q1) edge[bend left] node [below left]{$e^{10}$} (q10)
   (q10) edge node [above]{$e^{11}$} (q11)
   (q11) edge node [above]{$e^{12}$} (q12)
   (q12) edge node [above]{$e^{13}$} (q13)
   (q13) edge[bend left] node [above right]{$e^{14}$} (q6)
   
   (q2) edge[bend right, densely dashed] node [above left]{$e^{15}$} (q14)
   (q14) edge[bend right, densely dashed] node [ right]{$e^{16}$} (q7)
   (q4) edge[bend right, densely dashed] node [ right]{$e^{17}$} (q15)
   (q15) edge[bend right, densely dashed] node [ right]{$e^{18}$} (q9)
   (q12) edge[bend left=10, densely dashed] node [ above]{$e^{19}$} (q16)
   (q16) edge[bend left=10, densely dashed] node [ above right]{$e^{20}$} (q9)
   (q12) edge[bend left=10, densely dashed] node [ above right]{$e^{21}$} (q17)
   (q17) edge[bend left, densely dashed] node [ above right]{$e^{22}$} (q9);
            \end{tikzpicture}
        \end{adjustbox}
    \end{center}
    \caption{Original Automaton ($\mathcal{A}^{O}$ solid line) and Redundant Automaton ($\mathcal{A}^{R}$ dashed line) for a manufacturing layout.}
    \label{figure-sample PTA}
\end{figure}
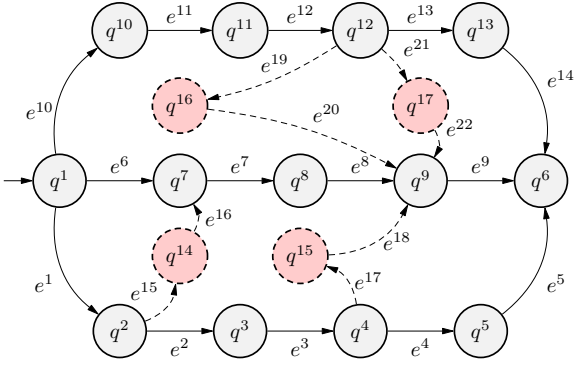
\begin{table}[t]
    \centering
    \renewcommand{\arraystretch}{1.2}
    \caption{List of the physical locations, costs, risk factors, and centralities of System 1 and System 2 in Figure \ref{figure-sys comparison}.}
    \label{tab:list of locations for systems}
    \begin{tabular}{|c|l|c|c|c|}
         \hline
         \multirow{2}{*}{$q^{i}$}&
         \multirow{2}{*}{Location}&
         \multirow{2}{*}{$P_{i}$}&
         \multirow{2}{*}{$h_{i}$}&
         \multirow{2}{*}{Centrality($x_{i}$)} \\
         &&&&\\
         \hline
         $q^1$& In transit (AGV1)& 1 & 1 & 3\\
         \hline
         $q^2$& CNC1 entry buffer & 1 & 1 & 2\\
         \hline
         $q^3$& CNC1 & 1 & 1 & 1\\
         \hline
         $q^4$& CNC1 exit buffer& 1 & 1 & 2\\
         \hline
         $q^5$& In transit (AGV2) & 1 & 1 & 1\\
         \hline
         $q^6$& Target buffer & 1 & 1 & 0\\
         \hline
         $q^7$& Emergency buffer 1 & 1 & 1 & 1\\
         \hline
         $q^8$& CNC2 & 1 & 1 & 1\\
         \hline
         $q^9$& In transit (AGV3) & 1 & 1 & 1\\
         \hline
         $q^{10}$& Emergency buffer 2 & 1 & 1 & 1\\
         \hline
         $q^{11}$& CNC3 entry buffer 2 & 1 & 1 & 1\\
         \hline
         $q^{12}$& CNC3 & 1 & 1 & 3\\
         \hline
         $q^{13}$& CNC3 exit buffer & 1 & 1 & 1\\
         \hline
         $q^{14}$& In transit (AGV4) & 1 & 1 & 1\\
         \hline
         $q^{15}$& Emergency buffer 3 & 1 & 1 & 1\\
         \hline
         $q^{16}$& Emergency buffer 4 & 1 & 1 & 1\\
         \hline
         $q^{17}$& Emergency buffer 4 & 1 & 1 & 1\\
         \hline
    \end{tabular}
    \vspace{-4mm}
\end{table}

\par
As for Scenario 2, $q^{11}$ fails initially. The risk-averse PTA MPC chooses the bottom path given that it has two redundant paths and that the top path has failed. Therefore, the bottom path has comparatively lower risk whereas PTA MPC chooses the middle path because it has the lowest cost. However, later at $t\geq t_{1}$ and $t\geq t_{2}$ both controllers face state failure at $q^{5}$ and $q^{8}$. This failure stops PTA MPC and it returns {\fontfamily{qcr}\selectfont UNSAT} while risk-averse PTA MPC utilizes the redundant path and reaches the desired state through path $\alpha^{\ast}=\langle q^{1},q^{2},q^{3},q^{4},q^{15},q^{9},q^{6} \rangle$ with objective function value $V^{\ast}=10.33$. 
\par
Finally, in Scenario 3, the bottom path fails as $q^{5}$ fails initially. Therefore, risk-averse PTA MPC chooses the top path which has two redundant paths while PTA MPC again chooses the middle path. When failure occurs in $q^{8}$ and $q^{13}$ at $t\geq t_{1}$ and $t\geq t_{2}$, PTA MPC stops and returns {\fontfamily{qcr}\selectfont UNSAT}, similar to Scenario 1. Risk-averse PTA MPC however, returns a pareto-frontier of two optimal paths $\alpha_{1}^{\ast}=\langle q^{1},q^{10},q^{11},q^{12},q^{16},q^{9},q^{6} \rangle$ and $\alpha_{2}^{\ast}=\langle q^{1},q^{10},q^{11},q^{12},q^{17},q^{9},q^{6} \rangle$ both with objective function value of $V^{\ast}=10.66$.
%\par
%The cost objective function value depends on the states that comprise a path. As occurrence of state failure is uncertain, the relative optimality of risk-averse PTA MPC compared to PTA MPC is also uncertain. While PTA MPC is more optimal on certain occasions, the risk-averse PTA MPC provides a more flexible solution which is robust to state failure. 

\section{Future Work and Conclusion}
\label{sec-conclusion}

In this work, we have proposed a risk-averse PTA-MPC that is flexible to sudden state failure. As part of this framework, we define an update operator and an emergency declaration signal that dynamically captures system disturbances and changes the PTA structure based on the manufacturing environment. We further defined a cost-based risk measure that allows the re-formulation of a  multi-objective constrained optimization into cost-dependent single-objective optimization. Finally, we developed risk-averse PTA MPC algorithm which returns the robust solution considering both risk and cost. The proposed framework was shown to allow the system to reach goal state in face of state failure when applied in a manufacturing system. It was also shown that under certain conditions, there exists a Pareto-frontier of solutions that can add extra flexibility in the system. 
This work can further be extended to other application areas, such as mass-production, supply-chain, and air-traffic control, where the objective of PTA MPC dynamic decision-making involves both cost and risk.
\par
For future work, intelligent scheduling of risk-factors can be investigated to ensure uninterrupted operation under time-varying demand for isolation, deviation or usage of a particular state. Scheduling EDS signals manually, stochastically or sensitive to number of state-occupation is another promising topic with applications in manufacturing under stochastically-uncertain or service-life dependent failures.
\bibliographystyle{IEEEtran}
% \printbibliography
\bibliography{main}

\begin{thebibliography}{10}
\providecommand{\url}[1]{#1}
\csname url@samestyle\endcsname
\providecommand{\newblock}{\relax}
\providecommand{\bibinfo}[2]{#2}
\providecommand{\BIBentrySTDinterwordspacing}{\spaceskip=0pt\relax}
\providecommand{\BIBentryALTinterwordstretchfactor}{4}
\providecommand{\BIBentryALTinterwordspacing}{\spaceskip=\fontdimen2\font plus
\BIBentryALTinterwordstretchfactor\fontdimen3\font minus
  \fontdimen4\font\relax}
\providecommand{\BIBforeignlanguage}[2]{{%
\expandafter\ifx\csname l@#1\endcsname\relax
\typeout{** WARNING: IEEEtran.bst: No hyphenation pattern has been}%
\typeout{** loaded for the language `#1'. Using the pattern for}%
\typeout{** the default language instead.}%
\else
\language=\csname l@#1\endcsname
\fi
#2}}
\providecommand{\BIBdecl}{\relax}
\BIBdecl

\bibitem{wang_current_2015}
\BIBentryALTinterwordspacing
L.~Wang, M.~Törngren, and M.~Onori, ``Current status and advancement of
  cyber-physical systems in manufacturing,'' vol.~37, pp. 517--527. [Online].
  Available:
  \url{https://linkinghub.elsevier.com/retrieve/pii/S0278612515000400}
\BIBentrySTDinterwordspacing

\bibitem{acatech_-_national_academy_of_science_and_engineering_2011_cyber-physical_2011}
\BIBentryALTinterwordspacing
``Cyber-physical systems.'' [Online]. Available:
  \url{http://link.springer.com/10.1007/978-3-642-29090-9}
\BIBentrySTDinterwordspacing

\bibitem{cassandras_introduction_2021}
C.~G. Cassandras and S.~Lafortune, \emph{Introduction to discrete event
  systems}, third edition~ed.\hskip 1em plus 0.5em minus 0.4em\relax Springer.

\bibitem{behrmann2005optimal}
G.~Behrmann, K.~G. Larsen, and J.~I. Rasmussen, ``Optimal scheduling using
  priced timed automata,'' \emph{ACM SIGMETRICS Performance Evaluation Review},
  vol.~32, no.~4, pp. 34--40, 2005.

\bibitem{kovalenko_priced_2020}
\BIBentryALTinterwordspacing
I.~Kovalenko, D.~Tilbury, and K.~Barton, ``Priced timed automata models for
  control of intelligent product agents in manufacturing systems,'' vol.~53,
  no.~4, pp. 136--142. [Online]. Available:
  \url{https://linkinghub.elsevier.com/retrieve/pii/S240589632100104X}
\BIBentrySTDinterwordspacing

\bibitem{cassandras_sensitivity_2008}
\BIBentryALTinterwordspacing
``Sensitivity analysis and concurrent estimation,'' in \emph{Introduction to
  Discrete Event Systems}, C.~G. Cassandras and S.~Lafortune, Eds.\hskip 1em
  plus 0.5em minus 0.4em\relax Springer {US}, pp. 617--740. [Online].
  Available: \url{http://link.springer.com/10.1007/978-0-387-68612-7_11}
\BIBentrySTDinterwordspacing

\bibitem{balta_model_2022}
\BIBentryALTinterwordspacing
E.~C. Balta, I.~Kovalenko, I.~A. Spiegel, D.~M. Tilbury, and K.~Barton, ``Model
  predictive control of priced timed automata encoded with first-order logic,''
  vol.~30, no.~1, pp. 352--359. [Online]. Available:
  \url{https://ieeexplore.ieee.org/document/9354443/}
\BIBentrySTDinterwordspacing

\bibitem{kovalenko_cooperative_2022}
\BIBentryALTinterwordspacing
I.~Kovalenko, E.~C. Balta, D.~M. Tilbury, and K.~Barton, ``Cooperative product
  agents to improve manufacturing system flexibility: A model-based decision
  framework,'' pp. 1--18. [Online]. Available:
  \url{https://ieeexplore.ieee.org/document/9737280/}
\BIBentrySTDinterwordspacing

\bibitem{hutchison_priced_2005}
\BIBentryALTinterwordspacing
G.~Behrmann, K.~G. Larsen, and J.~I. Rasmussen, ``Priced timed automata:
  Algorithms and applications,'' in \emph{Formal Methods for Components and
  Objects}, F.~S. de~Boer, M.~M. Bonsangue, S.~Graf, and W.-P. de~Roever,
  Eds.\hskip 1em plus 0.5em minus 0.4em\relax Springer Berlin Heidelberg, vol.
  3657, pp. 162--182, series Title: Lecture Notes in Computer Science.
  [Online]. Available: \url{http://link.springer.com/10.1007/11561163_8}
\BIBentrySTDinterwordspacing

\bibitem{behrmann_scheduling_2005}
\BIBentryALTinterwordspacing
G.~Behrmann, E.~Brinksma, M.~Hendriks, and A.~Mader, ``{SCHEDULING} {LACQUER}
  {PRODUCTION} {BY} {REACHABILITY} {ANALYSIS} – a {CASE} {STUDY},'' vol.~38,
  no.~1, pp. 50--55. [Online]. Available:
  \url{https://linkinghub.elsevier.com/retrieve/pii/S1474667016374456}
\BIBentrySTDinterwordspacing

\bibitem{hekmatnejad_task_2019}
\BIBentryALTinterwordspacing
M.~Hekmatnejad, G.~Pedrielli, and G.~Fainekos, ``Task scheduling with nonlinear
  costs using {SMT} solvers,'' in \emph{2019 {IEEE} 15th International
  Conference on Automation Science and Engineering ({CASE})}.\hskip 1em plus
  0.5em minus 0.4em\relax {IEEE}, pp. 183--188. [Online]. Available:
  \url{https://ieeexplore.ieee.org/document/8843048/}
\BIBentrySTDinterwordspacing

\bibitem{goos_as_2001}
\BIBentryALTinterwordspacing
K.~Larsen, G.~Behrmann, E.~Brinksma, A.~Fehnker, T.~Hune, P.~Pettersson, and
  J.~Romijn, ``As cheap as possible: Effcient cost-optimal reachability for
  priced timed automata,'' in \emph{Computer Aided Verification}, G.~Berry,
  H.~Comon, and A.~Finkel, Eds.\hskip 1em plus 0.5em minus 0.4em\relax Springer
  Berlin Heidelberg, vol. 2102, pp. 493--505, series Title: Lecture Notes in
  Computer Science. [Online]. Available:
  \url{http://link.springer.com/10.1007/3-540-44585-4_47}
\BIBentrySTDinterwordspacing

\bibitem{tilbury_cyber-physical_2019}
\BIBentryALTinterwordspacing
D.~M. Tilbury, ``Cyber-physical manufacturing systems,'' vol.~2, no.~1, pp.
  427--443. [Online]. Available:
  \url{https://www.annualreviews.org/doi/10.1146/annurev-control-053018-023652}
\BIBentrySTDinterwordspacing

\bibitem{panetto_challenges_2019}
\BIBentryALTinterwordspacing
H.~Panetto, B.~Iung, D.~Ivanov, G.~Weichhart, and X.~Wang, ``Challenges for the
  cyber-physical manufacturing enterprises of the future,'' vol.~47, pp.
  200--213. [Online]. Available:
  \url{https://linkinghub.elsevier.com/retrieve/pii/S1367578818302086}
\BIBentrySTDinterwordspacing

\bibitem{gray2020industrial}
M.~Gray-Hawkins and G.~L{\u{a}}z{\u{a}}roiu, ``Industrial artificial
  intelligence, sustainable product lifecycle management, and internet of
  things sensing networks in cyber-physical smart manufacturing systems,''
  \emph{Journal of Self-Governance and Management Economics}, vol.~8, no.~4,
  pp. 19--28, 2020.

\bibitem{ionescu2020digital}
L.~Ionescu \emph{et~al.}, ``Digital data aggregation, analysis, and
  infrastructures in fintech operations,'' \emph{Review of Contemporary
  Philosophy}, no.~19, pp. 92--98, 2020.

\bibitem{tarragona_systematic_2021}
\BIBentryALTinterwordspacing
J.~Tarragona, A.~L. Pisello, C.~Fernández, A.~de~Gracia, and L.~F. Cabeza,
  ``Systematic review on model predictive control strategies applied to active
  thermal energy storage systems,'' vol. 149, p. 111385. [Online]. Available:
  \url{https://linkinghub.elsevier.com/retrieve/pii/S1364032121006705}
\BIBentrySTDinterwordspacing

\bibitem{rawlings_model_2009}
J.~B. Rawlings and D.~Q. Mayne, \emph{Model predictive control: theory and
  design}, 1st~ed.\hskip 1em plus 0.5em minus 0.4em\relax Nob Hill Publ.

\bibitem{schwenzer_review_2021}
\BIBentryALTinterwordspacing
M.~Schwenzer, M.~Ay, T.~Bergs, and D.~Abel, ``Review on model predictive
  control: an engineering perspective,'' vol. 117, no.~5, pp. 1327--1349.
  [Online]. Available:
  \url{https://link.springer.com/10.1007/s00170-021-07682-3}
\BIBentrySTDinterwordspacing

\bibitem{schwenzer_review_2021-1}
\BIBentryALTinterwordspacing
------, ``Review on model predictive control: an engineering perspective,''
  vol. 117, no.~5, pp. 1327--1349. [Online]. Available:
  \url{https://link.springer.com/10.1007/s00170-021-07682-3}
\BIBentrySTDinterwordspacing

\bibitem{noauthor_tutorial_2000}
\BIBentryALTinterwordspacing
``Tutorial overview of model predictive control,'' vol.~20, no.~3, pp. 38--52.
  [Online]. Available: \url{https://ieeexplore.ieee.org/document/845037/}
\BIBentrySTDinterwordspacing

\bibitem{li_resource_2023}
\BIBentryALTinterwordspacing
A.~Li and J.~Sun, ``Resource limited event-triggered model predictive control
  for continuous-time nonlinear systems based on first-order hold,'' vol.~47,
  p. 101273. [Online]. Available:
  \url{https://linkinghub.elsevier.com/retrieve/pii/S1751570X2200070X}
\BIBentrySTDinterwordspacing

\bibitem{newman_networks_2018}
\BIBentryALTinterwordspacing
M.~Newman, \emph{Networks}.\hskip 1em plus 0.5em minus 0.4em\relax Oxford
  University Press, vol.~1. [Online]. Available:
  \url{https://oxford.universitypressscholarship.com/view/10.1093/oso/9780198805090.001.0001/oso-9780198805090}
\BIBentrySTDinterwordspacing

\bibitem{bjorner_z_nodate}
\BIBentryALTinterwordspacing
N.~Bjorner and A.-D. Phan, ``Vz maximal satisfaction with z3,'' pp. 1--9.
  [Online]. Available: \url{https://easychair.org/publications/paper/xbn}
\BIBentrySTDinterwordspacing

\bibitem{kovalenko_model-based_2019}
\BIBentryALTinterwordspacing
I.~Kovalenko, D.~Tilbury, and K.~Barton, ``\BIBforeignlanguage{en}{The
  model-based product agent: {A} control oriented architecture for intelligent
  products in multi-agent manufacturing systems},''
  \emph{\BIBforeignlanguage{en}{Control Engineering Practice}}, vol.~86, pp.
  105--117, May 2019. [Online]. Available:
  \url{https://linkinghub.elsevier.com/retrieve/pii/S0967066118305057}
\BIBentrySTDinterwordspacing

\bibitem{ocker_framework_2019}
\BIBentryALTinterwordspacing
F.~Ocker, I.~Kovalenko, K.~Barton, D.~Tilbury, and B.~Vogel-Heuser,
  ``\BIBforeignlanguage{en}{A {Framework} for {Automatic} {Initialization} of
  {Multi}-{Agent} {Production} {Systems} {Using} {Semantic} {Web}
  {Technologies}},'' \emph{\BIBforeignlanguage{en}{IEEE Robotics and Automation
  Letters}}, vol.~4, no.~4, pp. 4330--4337, Oct. 2019. [Online]. Available:
  \url{https://ieeexplore.ieee.org/document/8779665/}
\BIBentrySTDinterwordspacing

\end{thebibliography}

\end{document}